\documentclass[aps, reprint, nofootinbib]{revtex4-1}
\usepackage{amsmath, latexsym, amssymb, hyperref, graphicx, color, slashed}
\definecolor{nicered}{rgb}{0.7,0.1,0.1}
\definecolor{nicegreen}{rgb}{0.1,0.5,0.1}
\hypersetup{colorlinks, citecolor=nicegreen,linkcolor=nicered}
\begin{document}

\newcommand{\Sec}[1]{ \medskip \noindent {\sl \bfseries #1}}
\newcommand{\subsec}[1]{ \medskip \noindent {\sl \bfseries #1}}
\newcommand{\Par}[1]{ \medskip \noindent {\em #1}}

\title{Natural Philosophy versus Philosophy of Naturalness $^{\S}$}

\author{Goran Senjanovi\'c  }

\affiliation{International Centre for Theoretical Physics, Strada Costiera 11, I-34151 Trieste, Italy}

\affiliation{Tsung-Dao Lee Institute \& Department of Physics and Astronomy, SKLPPC, Shanghai Jiao Tong
University, 800 Dongchuan Rd., Minhang, Shanghai 200240, China}

\begin{abstract}

I reflect on some of the basic aspects of present day Beyond the Standard Model particle physics, focusing mostly on the issues of naturalness, in particular on the so-called hierarchy problem. To all of us, physics as natural science emerged with Galileo and Newton, and led to centuries of unparalleled success in explaining and often predicting new phenomena of nature. I argue here that the long standing obsession with the hierarchy problem as a guiding principle for the future of our field has had the tragic consequence of deviating high energy physics from its origins as  natural philosophy, and turning it into a philosophy of naturalness.  
\end{abstract}


\maketitle

{
 \renewcommand{\thefootnote}%
   {\fnsymbol{footnote}}
 \footnotetext[4]{Based on the closing talk of the International conference 'LHC Days in Split 2018', September 2018, Split, Croatia}
}

\subsection*{Disclaimer} \label{disclaimer}

   If the reader does not believe that physics is a rigorous science based on facts of nature - which I will call natural philosophy in this essay on the core of the beyond the Standard Model physics - no argument, no matter how logical and/or physical, could help in my experience to make him change his mind. It is unlikely that such a reader would profit from this write-up.
   My own prejudice is that it is necessary, while of course not sufficient, that you believe that physics should be based on predictive theories of natural phenomena in order that these reflections make sense to you. If you do believe that physics is a natural philosophy but you find yourself caught on the bandwagon of naturalness, please take your time to reflect over the arguments presented here. You can rest assured that I will do the same.

\section{Prologue} \label{Prologue}

    About a year and a half ago, I was asked by the organizers of the {\it LHC Days in Split 2018} conference to give a kind of vision talk, an overview of the present day situation in our field and to offer some thoughts on what the future could bring. Since it makes no sense to pretend to know what lies ahead, I tried to look back into what has been achieved in the past. I attempted in particular to assess what happened to our field in the last thirty five years, after the $W$ and $Z$ boson were discovered. This epoch was characterized by the great success of neutrino physics, culminated with the proof of non-vanishing neutrino mass through neutrino oscillations. And yet, the dominant theme of the Beyond Standard Model physics, strangely enough, was not the search for the fundamental theory behind neutrino mass but rather the obsession with the so-called hierarchy problem behind Higgs boson mass. Most spokespersons would argue that the unnatural smallness of Higgs mass compared to the Planck scale or the grand unification scale requires the new physics to lie at the TeV energies - if not below. The fact that the (weak scale)/(Planck scale) mass ratio is so small became the infamous hierarchy problem, since there is no protective symmetry behind. This was the prime example of the naturalness issues that started to drive the research into asking why the numbers are what they are, and how to protect them when they are very small. 
  
     The hierarchy problem was argued to be cured by the beautiful possibility of low energy supersymmetry. I myself was caught in the whirlpool of this sentiment and fell in love with it. In 1981, working on the high precision analysis of low energy supersymmetry unification, inspired by the work of Dimopoulos et al~\cite{Dimopoulos:1981yj}, Bill Marciano and I suggested that the predicted value of the weak mixing angle - then claimed wrongly to be smaller - required the top quark mass to lie around 200 GeV~\cite{Marciano:1981un}. That was a blasphemy in that era - everybody 'knew' that top mass had to be close to the bottom one - but when it turned out correct (and meanwhile the weak mixing angle was found at LEP to be $\theta_W = 0.23$ in accord with the low energy supersymmetry prediction), I as many others was hooked on it. Over the years though I slowly started to lose my faith, not so much due to supersymmetry not being seen - after all, even if it lies above TeV, it still protects us from the peril of the high scales - but because the Minimal Supersymmetric Standard Model turned out to be a collection of myriads of different models depending on what you assume about the masses and mixings of the super-partners. Moreover, why in the world should the super-partners all lie at one and the same scale in view of the enormous spread of the SM particle masses? 
  Models can be constructed that achieve that, but models can be constructed that achieve basically any wishful thinking.
     
     Equally important, at the beginning of the century it finally became  universally accepted that neutrinos are massive and the seesaw mechanism emerged as the main explanation behind the  smallnes of its mass~\cite{Minkowski:1977sc}. In the context of grand unification, the seesaw fits nicely with an intermediate scale - the same scale that fits nicely with neutrino mass - in the SO(10) theory, and no supersymmetry is needed for the sake of the unification of couplings. So the only true motivation for low energy supersymmetry remained the hierarchy issue of the Higgs mass.
     
     On the other hand, with time the SM turned into a high precision theory, and thus clearly it had no problem of inconsistency of any kind - so why in the world would it have to be modified at the TeV scale? I started to feel more and more that the explanation for the Higgs boson lightness could have to do with whatever could be the consistency requirement of some fundamental high energy physics, or some deep, presently unknown, principle. At the same time, low energy supersymmetry started to dominate the BSM model {\it ad nauseam}. To appreciate the degree of supersymmetry fever, think of this incredible fact -  any deviation from the SM is called exotics by the LHC workgroups - even the attempts to understand the origin of neutrino mass through minimal extensions of the SM. Not low energy supersymmetry though, in spite of the total lack of physical, phenomenological motivation.

       Meanwhile another important development took place - an explicit realization of the possibility of the TeV cut-off of field theory in the context of large extra dimensions, which implied that there may not even be high energy field theory~\cite{ArkaniHamed:1998rs}. Still, in spite of my original enthusiasm for this striking possibility, I again felt that there was nothing urgent about the hierarchy problem, but  rather that  BSM physics should be based on physical, phenomenological arguments. The field stayed firmly on its grounds though, and I started to worry that this obsession with naturalness of physics parameters was taking us in a wrong direction. After all, the great success of post-Galilean physics was precisely due to it having become a natural philosophy, as called eloquently by Newton (and emphasised lot by Weinberg in modern times) and practiced by more than four centuries - a precise and predictive study of natural phenomena driven by phenomenological observations and philosophical implications, but never just the latter without the former. 

  {\bf In short, high energy physics, instead of being natural philosophy, was being transformed into a philosophy of naturalness.}
  
   It has been some years that I wanted to write an essay along these lines and now I am additionally inspired by a recent paper by Dvali~\cite{Dvali:2019mhn}, which I feel  vindicates my point of view.  Let me be clear here: it is not that I feel that the hierarchy question is not fundamentally important - I only claim that it is not urgent, in a sense of requiring a change of low energy physics. Dvali's work provides an interesting and profound scenario of how this could come about, by means of a consistency argument (based on his early work with Vilenkin~\cite{Dvali:2003br})  that could let us live with the low value of the weak scale, even if the new physics lies at the arbitrarily large scale. In simple terms he argues that the vacuum in which live, with the observed value of the Higgs field much smaller than the Planck scale, may not be problem at all since it could be a product of inflationary evolution.
I personally would not say that this provides the solution, but that it shows that the problem was never there. For a believer in physics as natural philosophy, a solution to a problem by definition requires well defined observable new phenomena. One may argue that this is simply a question of semantics, but semantics is important since it influences the way we think and work.
   Dvali also points out what many of us felt for a long time - that the naturalness program failed badly in the case of the cosmological constant, an additional argument against the obsession with naturalness.
   
   In any case, I feel that time is ripe to discuss the essence of what I believe is wrong with the hierarchy problem bandwagon. I start with a general, albeit brief, discussion of naturalness and then move to the specific issue of the Higgs mass.

 \section{Naturalness in a nutshell} \label{naturalness}
 
    The concept of naturalness has been nicely reviewed in~\cite{Dvali:2019mhn}. Let me phrase it here once again for the sake of completeness. Take a physical parameter
    $\theta$ which we can divide in the classical, tree-level, part and the quantum part of radiative corrections 
  \begin{equation}\label{theta}
\theta = \theta_{tree} + \theta_{loop}.
 \end{equation}
The parameter $\theta$ may be dimension-full (a particle mass, say) or dimension-less (a coupling). If it happens that 
  \begin{equation}\label{loop}
 \theta_{loop} \gg \theta_{exp},.
 \end{equation}
where $\theta_{exp}$ is an experimental value (or bound), we say that the parameter 
$\theta$ is unnaturally small. It then requires the fine-tuned cancellation between 
$\theta_{tree}$ and $\theta_{loop}$, something that most of us would consider rather unappealing. This has been named a naturalness, or fine-tuning, problem. When 
$\theta$ is dimension-full it is called a hierarchy problem, since it requires a hierarchy of scales.

 Notice however that this is not a problem - it can always be done at the needed precision and at a chosen order of perturbation theory - but it is certainly a puzzling question. As I will argue below, calling it a problem obscures the issue and creates confusion. 
There is definitely no problem of naturalness when a protective symmetry keeps $\theta_{loop}$ small in perturbation theory, such as say chiral symmetry in the limit of vanishing fermion mass.  
 
 The case of enhanced symmetry in the limit $\theta_{tree} = 0$ for a physical parameter $\theta$ has been called technical naturalness, or naturalness a la 't Hooft~\cite{tHooft:1979rat}. No need for fine tuning in such a case, which provides a welcome relief. I wish to emphazise here that there is a truly natural, intuitive  definition of naturalness, something that happens in our daily experience.  Small fermion masses are technically
 natural but equally important they should be considered natural precisely since most of them are small. Most fermions are light compared to the electro-weak scale responsible for their masses, which is tantamount to small Yukawa couplings. And yet, more often than not, a proponent of a new model will argue that the natural values for new Yukawa couplings ought to be of order one. Even worse, in many instances simple models are made baroque upon learning that some couplings have to be very small.
 
    There is much to be said about technical vs natural naturalness, but suffice it to say that the naturalness concept is often abused in demanding that a technically natural small parameter $\theta$ vanish by ad-hoc use of a global symmetry, instead of studying the experimental limits on $\theta$. In the case of protective symmetry, zero is not a special point and thus any small value for $\theta$ is equally acceptable. After all, zero is smaller than small.
    
   It would be tautology to say that the proton must be absolutely stable due to the baryon number symmetry which protects its longevity. It would be equally tautological to argue that proton-neutron mass difference ought to vanish due to the isospin symmetry which protects its smallness. This I am sure is obvious to everybody, and yet such arguments are constantly used in literature. The prime example is the celebrated $Z_2$ symmetry~\cite{Glashow:1976nt} in the case of two Higgs doublets which guarantees neutral current flavor conservation at the tree-level in all of the parameter space. This is a strange demand - after all the GIM mechanism works only for a small charm quark mass. The $Z_2$ extreme limit was justifiably
 criticized in~\cite{Gogberashvili:1991ws}, since there is nothing special about the point of exact global symmetry, but is pursued to this day and used over and over.
    
    Let us however in what follows concentrate on a particular case of a dimension-full parameter, mass. As we said a fermion mass is protected by the chiral symmetry. Similarly, a gauge boson mass is protected by a local gauge symmetry (which can actually be regained even for a massive gauge boson) so the issue is only relevant for a scalar mass.

\section{The hierarchy problem} \label{hierarchy}
    
    We are now  led to supersymmetry, since in this case there is no difference between scalars and fermions, and the same chiral symmetry that protects fermion mass also protects the scalar one. This is what makes supersymmetry so special from the hierarchy point of view and this is why one would love the scale of supersymmetry breaking in our world to be not too far from the Higgs mass scale, or the $W$ mass. 
    
    But how far is not too far? Notice how this becomes immediately an aesthetic or emotional issue as to when fine tuning is too much - a personal feeling more than the scientific criterium. In any case, the fine tuning - if it is there - persists even with supersymmetry, no matter how low its breaking scale may end up being. 
    
    After all, in the SM there is no fine tuning whatsoever, not unless you bring up explicitly a new high energy scale, such as in the case of grand unification. In that case, the SM Higgs doublet cannot live alone, since it must be embedded into a representation of a GUT gauge group. A celebrated example is provided by the minimal SU(5) theory where the SM Higgs lives in a 5-dimensional representation $5_H$, consisting of the SM doublet D and a new heavy colour triple, weak singlet scalar T. The latter field must be heavy and live practically at the GUT scale $M_{GUT}$ since it mediates proton decay. The masses of D and T gets split up by the vev of the adjoint Higgs $24_H$ needed for the GUT scale symmetry breaking. In the process the D and T masses become (up to irrelevant Clebsch-Gordan coefficients)
  \begin{equation}\label{Dmass}
  m_D^2 = m_5^2 - M_{GUT}^2,  \,\,\,\,\, m_T^2 = m_5^2 + M_{GUT}^2
 \end{equation}
where $m_5$ is the original symmetric mass term of the $5_H$ field. Since $m_T$ must be not much smaller than  $M_{GUT}$ ($T$ mediates proton decay), this implies 
$m_5 \simeq M_{GUT}$ to an astonishing precision. This is the infamous doublet-triplet (D-T) splitting problem, an explicit example of mandatory fine-tuning. In the context of grand unification, where a new huge mass scale is a must, the issue of fine-tuning is clearly very painful.

What happens in the supersymmetric extension of the theory? The answer unfortunately is the same painful fine-tuning and the reason is obvious. Once again one must split doublet and triplet multiplet masses, and once again this is achieved by having the $24_H$ acquire a huge vev, of order $M_{GUT}$ - hence precisely the same problem all over. Supersymmetry {\it per se} offers no help - all it does is to keep the fine-tuning stable as long as the scale of supersymmetry breaking is not far from the electro-weak one. 

This to me is a sufficient reason to be cautious about supersymmetry. True, it predicted the gauge coupling unification, but this can be also achieved naturally without it, with a large seesaw scale. In the context of the $SO(10)$ grand unified theory, an intermediate seesaw scale 
 provides sufficiently large neutrino mass and at the same time, predicts the unification of the SM gauge couplings.
The point is that low energy supersymmetry does not answer the mystery of the SM scale being so much below the GUT scale. Of course, one can employ group theoretic tricks to achieve the D -T splitting without fine tuning, but it is typically done with a change at $M_{GUT}$, a change that does not affect low energy physics. A particular approach stands out in that sense, Higgs as a pseudo-Goldstone boson of an accidental symmetry~\cite{Inoue:1985cw}. But even this appealing idea ends up having no impact on the low energy world that can be probed in any foreseeable future. True, it requires low energy supersymmetry to protect the Higgs from going to the GUT scale, but the pseudo-Goldstone nature of the Higgs is hidden from the low energy observer. 

An important comment. Neither the desire to protect the $D-T$ splitting from loops effects, once fine-tuned, nor the gauge coupling unification can give a quantitative prediction for the scale of supersymmetry breaking (defined here as the difference in masses between particles and super-partners). Unlike what is often claimed, this scale could easily be above the LHC reach without anything being lost. It is not even clear that the super-partner's masses ought to lie at one and the same scale, even if one worries about the Higgs mass protection, since lighter generations could have much heavier super-partners without any visible effect. Just as it should have never be claimed that supersymmetry ought to be observed at the LHC (or LEP as it was argued in the eighties), it is senseless to say that it is being ruled out by not being seen at the LHC. The low energy supersymmetry is as alive or as dead as it ever was.

In any case, I, for one, remain skeptical of the claim that new observable physics must stem from the solution to this profound issue. We have no clue of a more fundamental theory of nature, and it could well be that the consistency mechanism behind it may unravel this mystery without an impact on the TeV, as argued by Dvali~\cite{Dvali:2019mhn}.

  Before I move to discuss the physical motivation for new physics at the energies accessible to the LHC or next hadron collider, a few words  more about grand unification. Needless to say, it remains a beautiful, deeply motivated, extension of the SM with profound consequences: magnetic monopoles and proton decay. I myself have spent years working on it, both ordinary and supersymmetric versions, trying hard (as many others) to come up with a predictive theory of proton decay branching ratios. Except for the minimal, original $SU(5)$ Georgi-Glashow theory,~\cite{Georgi:1974sy} ruled out by experiment, no other model is truly predictive due to large threshold effects from many new states around the GUT scale, a discouraging fact for a believer in predictivity as an essential criterium in a search for new physics. 
  
  It may be worth mentioning that a minimal extension~\cite{Bajc:2006ia}  of the minimal $SU(5)$ that can account for gauge coupling unification and neutrino mass predicts new particles at the TeV energies, but still, the model fails to make clear statements about proton decay channels. In what follows I will not talk about grand unification any more, not because I do not find it fundamentally important but rather since I am here after the guidance for new physics at the TeV energies - and grand unification does not make a convincing argument in its favor. 
      
  \section{If not hierarchy, then what?} \label{what}  
  
  So if physics is natural philosophy, as we should all believe, and not a philosophy of naturalness - what could be the new physics at the TeV scale being probed by the LHC? Instead of a hierarchy problem, what could be a motivation for the BSM physics? The answer is obvious. The only true failure of the SM is the prediction of the vanishing neutrino mass - clearly the origin of neutrino mass ought to be the door to new physics. The question then is why should we hope that the LHC or the next generation hadron collider could open that door? The answer lies in the physically attractive possibility of neutrino being Majorana particle. Since it is only its vanishing charge that separates neutrino from its electro-weak sibling the electron, it is natural that it may be behind the smallness of its mass. And sure enough, it is precisely the Majorana nature of neutrino that leads to the seesaw mechanism. Now, one often argues that the natural scale of seesaw is large, close to $M_{GUT}$ since then the Dirac Yukawa coupling becomes of order one. But is this true?
  
  After all, large scales bring in the same hierarchy issue that one argues to be a real problem. It is surely more technically natural to have a lower scale and small Yukawa couplings, in view of their self-protection. Moreover, the SM is all about small Yukawa couplings of charged fermions, so it is more natural for Dirac Yukawas to be small, in the natural sense of the word natural. 
  
  Notice however that once again we have fallen into the trap of being philosophical about physical scales, once again we have let the naturalness criterium tell us where new physics ought to lie. This is wrong. The scale issue ought to be dictated by phenomenological considerations.
  So what about phenomenological arguments? The answer lies in the text-book consequence of the Majorana nature of neutrino: neutrinoless double beta decay. To get a feel for what is expected, recall that the measure of this process in the case of being mediated by neutrino Majorana mass is given by
  \begin{equation}\label{nuamp}
{\cal A}_\nu \simeq G_F^2 \frac{m_\nu}{p^2},
 \end{equation}
  where $p$ is a measure of neutrino virtuality and typically on order of 100 MeV for nuclei of experimental interest. Since neutrinoless double beta is a d=9 six fermion process, the coefficient above must be cut-off by the fifth power of mass - as it is.
  
  The present and near future experiments are probing neutrino Majorana mass to about $m_\nu \simeq 10^{-1} eV$. In any case, from $m_\nu \lesssim 1 eV$, one has
  \begin{equation}\label{nuest}
{\cal A}_\nu \lesssim 10^{-17} GeV^{-5}.
 \end{equation}

    Now imagine that tomorrow this process is observed. What would that imply? One possibility of course is a non-vanishing neutrino Majorana mass, but it is far from being unique since the SM with Majorana neutrino needs an UV completion. But then the new physics may as well be the culprit behind this process, and moreover, it would have to be if the electrons that come out of the neutrinoless double beta decay were for example right-handed. Or if it was established that the neutrino mass hierarchy is what we call normal, since in this case the neutrino contribution is expected to be too small to cause an effect observable today (or tomorrow). The fact that  neutrinoless double beta decay is not really a probe of neutrino mass has been argued already sixty years ago by Feinberg and Goldhaber and echoed by Pontecorvo~\cite{maurice}, and yet to this day often ignored. In our seesaw paper~\cite{Minkowski:1977sc} Mohapatra and I have even provided an explicit realization in the form of the RH neutrino and the RH gauge boson as mediators of this process.
    
    In any case, whatever new physics may lie behind  neutrinoless double beta, it would have to be quite low in order to cause an observable effect. To see this, write its contribution in the effective form, for simplicity and illustration imagined with a single new mass scale
  \begin{equation}\label{nuest}
{\cal A}_{NP} \simeq \Lambda ^{-5}
 \end{equation}
      From \eqref{nuest}, one has in turn $\Lambda \geq TeV$. In other words, the cut-off for this process is on the order of few TeV, tailor-made for the LHC. And new physics is a direct lepton number violation at hadron colliders in a form of same sign lepton pairs accompanied by jets as argued by Keung and myself~\cite{Keung:1983uu} more than thirty years ago. It seemed almost science fiction at that time, but the LHC has made it an exciting reality, and today both CMS and ATLAS are actively pursuing it~\cite{Aaboud:2019wfg}. Let me not dwell on it, the interested reader should consult~\cite{Nemevsek:2018bbt}.

     In summary, if new physics were to cause neutrinoless double decay, it could lie tantalisingly close to the LHC energies. Hard to imagine a better motivation for observable new physics, suggested by pure phenomenological considerations - and yet it is rarely mentioned. Ironically, neutrinoless double beta decay could actually be a probe of the new physics behind neutrino Majorana mass and not the probe of the mass itself. This possibility would be even more exciting for it would open the door to lepton number violating processes at today's or near future hadron colliders.

     \section{Epilogue} \label{Epilogue}
        
        By now the reader has hopefully grasped the essence of my trouble with the present day atmosphere in our field, but let me make sure once again that I get my message across as clearly as possible. I have recently come back from a major conference and as usual I  heard speaker after speaker talk about the shortcomings and the problems of the SM: the hierarchy problem, the strong CP problem, the fermion mass problem and so on and so forth, all used as a guidance for the search of new physics. And yet none of these problems based on naturalness are problems at all, on the contrary they impose no constraint whatsoever on the parameter space of the SM. The only problem of the SM is a lack of problems. Even the strong CP violation imposes no constraint on its parameter space, strangely enough. 
        
        True, there is dark matter but we have no idea at all what it is and if it ends up being black holes of one type or another, it would not require new physics whatsoever. 
        
      Fortunately the SM did fail, and it failed loudly and clearly by predicting a vanishing neutrino mass. Arguably then, the most obvious and the only true phenomenological road to observable BSM physics ought to be the origin of neutrino mass.  I find it mind boggling that this is not universally shared. To me everything says then that 
     high energy physics is alive and well, we have a plethora of experiments sensitive to new physics scale that may be accessible if not to the LHC itself, at least to the future hadron collider. And it has nothing to do with the naturalness or hierarchy issues, issues that are more emotional or philosophical rather than phenomenological. Simply, once again, at the risk of boring the reader to death: neutrinoless double beta, the process that touches into the essence of up to now sacred law of lepton number conservation, if observed could well point out to new physics. And this new physics would have all the chance in the world to lie at the few TeV energies as I explained in detail above. 
     
     A particularly illustrative example of such new physics is the Left-Right Symmetric Model~\cite{Pati:1974yy} that led originally to the prediction of 
   non-vanishing neutrino mass. This theory started as an attempt to attribute maximal parity violation to its spontaneous symmetry breaking and over the years turned into the self-contained predictive theory of neutrino mass. In particular it can untangle the seesaw mechanism, allowing for the verifiable Higgs origin of neutrino mass, as originally shown with Miha Nemev\v{s}ek and Vladimir Tello~\cite{Nemevsek:2012iq}, analogous to the Higgs origin of charged fermion masses in the SM. This  remarkable result came as a surprise even to the authors (I can vouch for at least one of them) and in recent years Tello and I carried this program to the bitter end, see e.g.~\cite{Senjanovic:2019moe} for a recent discussion and references therein. It should be stressed - and it cannot be overstressed - that left-right symmetry is not used here in tautological sense by predicting that the parameters that violate it must vanish. Rather, it simply says that the dominant breaking is dynamical with resulting physical consequences. Also, left-right symmetry is not something chosen ad-hoc in order to control the unknown part of a parameter space, as is often done in model building. It is the first symmetry that the child sees and it shapes our understanding of the world, culminating with its breaking in weak interactions. Its fundamental role is similar to the one played by Lorentz symmetry or even more basically by the invariance under the laws of mechanics discovered by Galileo.~\footnote{I am grateful to G. Dvali for emphasising this point regarding the importance of left-right symmetry, the very reason for my original entusiasm.}

   Imposing such symmetries is surely justifiable as long as we do not claim that that they must be exact or eternal.

      In short, we can have phenomenological and experimental facts guide us to beyond the SM physics, as they did over and over in the past. All that is needed is a sense of discipline and rigour and  avoiding at all cost pessimistic proclamations of the end of phenomenologically based particle physics. These proclamations are the result of the incredibly optimistic feelings that all that is left to be done is to explain why numbers are what they are (and not what they are not). The history of physics is full of such  social movements that wanted to end physics and they have always been rather damaging, to put it mildly. Think of the bootstrap, Regge poles and such that were to represent the end of fundamental physics in the sixties and think of all the incredible progress that was being made precisely while the doom was preached: quarks with color and its gauging and the asymptotic freedom that led to QCD as a theory of strong interactions; the electro-weak gauge theory, the ideas of spontaneous symmetry breaking with Nambu-Goldstone and Higgs mechanisms, some of the greatest achievements in the history of physics. And incredibly enough, almost no attention was being paid to these developments when they were taking place, maybe precisely due to the prevailing atmosphere that there was nothing basic to be uncovered any more. How can you recognize a new development when you sustain that such a thing cannot exist?

 \subsection*{Acknowledgments}
  
    The author acknowledges a quarter of century of dialectic and often fierce debates with Gia Dvali on the hierarchy and naturalness issues discussed here. He is grateful to Charanjit Aulakh, Alessio Maiezza, Alejandra Melfo, Miha  Nemev\v{s}ek, Fabrizio Nesti, Vladimir Tello and Francesco Vissani for many fruitful discussions over the years on a number of issues discussed above. He also uses the opportunity to apologise to them for complaining too often against the bandwagon of naturalness (and other wagons) instead of writing this a long time ago. 
     Thanks are due to Gia Dvali, Alejandra Melfo and Vladimir Tello for their encouragement, 
  and help in improving the quality and clarity of my presentation. I am also grateful to Alessio Maiezza and Fabrizio Nesti for careful reading of the manuscript.
       
     I initiated this essay during the visit to the theory division of Fermilab last summer and  did most of the writing while visiting the particle physics division of Tsung-Dao Lee institute in Shanghai.  I am grateful to both institutions for their warm hospitality, in particular to Wai-Yee Keung and Stephen Parke for making me feel at home.


\begin{thebibliography}{99}

\bibitem{Dimopoulos:1981yj} 
  S.~Dimopoulos, S.~Raby and F.~Wilczek,
  ``Supersymmetry and the Scale of Unification,''
  Phys.\ Rev.\ D {\bf 24}, 1681 (1981).
  doi:10.1103/PhysRevD.24.1681

\bibitem{Marciano:1981un} 
  W.~J.~Marciano and G.~Senjanovi\'c,
  ``Predictions of Supersymmetric Grand Unified Theories,''
  Phys.\ Rev.\ D {\bf 25}, 3092 (1982).
  doi:10.1103/PhysRevD.25.3092
  
\bibitem{Minkowski:1977sc}
P.~Minkowski,
``Mu $\to$ E Gamma At A Rate Of One Out Of 1-Billion Muon Decays?,''
Phys.\ Lett.\ B {\bf 67} (1977) 421.


  R.~N.~Mohapatra and G.~Senjanovi\'c,
  ``Neutrino Mass and Spontaneous Parity Violation,''
  Phys.\ Rev.\ Lett.\  {\bf 44} (1980) 912.
  
  
  T.~Yanagida,
  ``Horizontal Symmetry And Masses Of Neutrinos,''
  Conf.\ Proc.\ C {\bf 7902131}, 95 (1979).

  S.~L.~Glashow,
  ``The Future of Elementary Particle Physics,''
  NATO Sci.\ Ser.\ B {\bf 61}, 687 (1980).
  
  

  M.~Gell-Mann, P.~Ramond and R.~Slansky,
  ``Complex Spinors and Unified Theories,''
  Conf.\ Proc.\ C {\bf 790927} (1979) 315
  [arXiv:1306.4669 [hep-th]].





\bibitem{ArkaniHamed:1998rs} 
  N.~Arkani-Hamed, S.~Dimopoulos and G.~R.~Dvali,
  ``The Hierarchy problem and new dimensions at a millimeter,''
  Phys.\ Lett.\ B {\bf 429}, 263 (1998)
  doi:10.1016/S0370-2693(98)00466-3
  [hep-ph/9803315].

\bibitem{Dvali:2019mhn} 
  G.~Dvali,
  ``Cosmological Relaxation of Higgs Mass Before and After LHC and Naturalness,''
  arXiv:1908.05984 [hep-ph].
  
\bibitem{Dvali:2003br} 
  G.~Dvali and A.~Vilenkin,
  ``Cosmic attractors and gauge hierarchy,''
  Phys.\ Rev.\ D {\bf 70}, 063501 (2004)
  doi:10.1103/PhysRevD.70.063501
  [hep-th/0304043].
  
\bibitem{tHooft:1979rat} 
  G.~'t Hooft,
  ``Naturalness, chiral symmetry, and spontaneous chiral symmetry breaking,''
  NATO Sci.\ Ser.\ B {\bf 59}, 135 (1980).

\bibitem{Glashow:1976nt} 
  S.~L.~Glashow and S.~Weinberg,
  ``Natural Conservation Laws for Neutral Currents,''
  Phys.\ Rev.\ D {\bf 15}, 1958 (1977).
  doi:10.1103/PhysRevD.15.1958
  
\bibitem{Gogberashvili:1991ws} 
  M.~Y.~Gogberashvili and G.~R.~Dvali,
  ``Hierarchy at Yukawa constants and K0 anti-K0, B0 anti-B0 oscillations in the model with two Higgs doublets,''
  Sov.\ J.\ Nucl.\ Phys.\  {\bf 53}, 491 (1991)
  [Yad.\ Fiz.\  {\bf 53}, 785 (1991)].

  L.~J.~Hall and S.~Weinberg,
  ``Flavor changing scalar interactions,''
  Phys.\ Rev.\ D {\bf 48}, R979 (1993)
  doi:10.1103/PhysRevD.48.R979
  [hep-ph/9303241].
  
\bibitem{Inoue:1985cw} 
  K.~Inoue, A.~Kakuto and H.~Takano,
  ``Higgs as (Pseudo)Goldstone Particles,''
  Prog.\ Theor.\ Phys.\  {\bf 75}, 664 (1986).
  doi:10.1143/PTP.75.664
  
  A.~A.~Anselm and A.~A.~Johansen,
  ``SUSY GUT with Automatic Doublet - Triplet Hierarchy,''
  Phys.\ Lett.\ B {\bf 200}, 331 (1988).
  doi:10.1016/0370-2693(88)90781-2
  
  Z.~G.~Berezhiani and G.~R.~Dvali,
  ``Possible solution of the hierarchy problem in supersymmetrical grand unification theories,''
  Bull.\ Lebedev Phys.\ Inst.\  {\bf 5} (1989) 55
   [Kratk.\ Soobshch.\ Fiz.\  {\bf 5} (1989) 42].
  
  
\bibitem{Georgi:1974sy} 
  H.~Georgi and S.~L.~Glashow,
  ``Unity of All Elementary Particle Forces,''
  Phys.\ Rev.\ Lett.\  {\bf 32}, 438 (1974).
  doi:10.1103/PhysRevLett.32.438
  
\bibitem{Bajc:2006ia} 
  B.~Bajc and G.~Senjanovi\'c,
  ``Seesaw at LHC,''
  JHEP {\bf 0708}, 014 (2007)
  doi:10.1088/1126-6708/2007/08/014
  [hep-ph/0612029].
  
  B.~Bajc, M.~Nemev\v{s}ek and G.~Senjanovi\'c,
  ``Probing seesaw at LHC,''
  Phys.\ Rev.\ D {\bf 76}, 055011 (2007)
  doi:10.1103/PhysRevD.76.055011
  [hep-ph/0703080].

    \bibitem{maurice}
	G.~Feinberg, M.~Goldhaber,
	Proc.\ Nat.\ Ac.\ Sci.\ USA {\bf 45} (1959) 1301.
	
  B.~Pontecorvo,
  ``Superweak interactions and double beta decay,''
 Phys.\ Lett.\  {\bf B26 } (1968)  630.


\bibitem{Mohapatra:1979ia}
  R.~N.~Mohapatra and G.~Senjanovi\'c,
  ``Neutrino Mass and Spontaneous Parity Violation,''
  Phys.\ Rev.\ Lett.\  {\bf 44} (1980) 912.
  
  
\bibitem{Keung:1983uu}
  W.~Y.~Keung and G.~Senjanovi\'c,
  ``Majorana Neutrinos And The Production Of The Right-Handed Charged Gauge
  Boson,''
  Phys.\ Rev.\ Lett.\  {\bf 50}, 1427 (1983).
  
    
\bibitem{Aaboud:2019wfg} 
  M.~Aaboud {\it et al.} [ATLAS Collaboration],
  ``Search for a right-handed gauge boson decaying into a high-momentum heavy neutrino and a charged lepton in $pp$ collisions with the ATLAS detector at $\sqrt{s}=13$ TeV,''
  Phys.\ Lett.\ B {\bf 798}, 134942 (2019)
  doi:10.1016/j.physletb.2019.134942
  [arXiv:1904.12679 [hep-ex]].
  
\bibitem{Nemevsek:2018bbt} 
  M.~Nemev\v{s}ek, F.~Nesti and G.~Popara,
  ``Keung-Senjanovi\'c process at the LHC: From lepton number violation to displaced vertices to invisible decays,''
  Phys.\ Rev.\ D {\bf 97}, no. 11, 115018 (2018)
  doi:10.1103/PhysRevD.97.115018
  [arXiv:1801.05813 [hep-ph]].


%
 %
\bibitem{Pati:1974yy} 
  J.~C.~Pati and A.~Salam,
  ``Lepton Number as the Fourth Color,''
  Phys.\ Rev.\ D {\bf 10}, 275 (1974)
  Erratum: [Phys.\ Rev.\ D {\bf 11}, 703 (1975)].
  doi:10.1103/PhysRevD.10.275, 10.1103/PhysRevD.11.703.2

  R.~N.~Mohapatra and J.~C.~Pati,
  ``A Natural Left-Right Symmetry,''
  Phys.\ Rev.\ D {\bf 11}, 2558 (1975).
  doi:10.1103/PhysRevD.11.2558
 
G.~Senjanovi\'c and R.~N.~Mohapatra,
``Exact Left-Right Symmetry And Spontaneous Violation Of Parity,''
Phys.\ Rev.\ D {\bf 12} (1975) 1502.


  G.~Senjanovi\'c,
  ``Spontaneous Breakdown of Parity in a Class of Gauge Theories,''
  Nucl.\ Phys.\ B {\bf 153}, 334 (1979).
  doi:10.1016/0550-3213(79)90604-7
  
\bibitem{Nemevsek:2012iq}
  M.~Nemev\v{s}ek, G.~Senjanovi\'c and V.~Tello,
  ``Connecting Dirac and Majorana Neutrino Mass Matrices in the Minimal Left-Right Symmetric Model,''
  Phys.\ Rev.\ Lett.\  {\bf 110} (2013) 15,  151802
  [arXiv:1211.2837 [hep-ph]].

Notice that the title in the published version is different (and much less reader friendly) from the original one in the arXiv (courtesy of PRL). Looking back, a title such as ``Higgs Origin of Majorana Neutrino Mass'' could have probably helped better to make the case.
  
\bibitem{Senjanovic:2019moe} 
  G.~Senjanovi\'c and V.~Tello,
  ``Parity and the origin of neutrino mass,''
  arXiv:1912.13060 [hep-ph].
  
I am grateful to G. Dvali for emphasising strongly this point regarding the importance of left-right symmetry, the very reason for my original interest and sustained dedication to its role in understanding weak interactions and neutrino mass issue.



\end{thebibliography}
\end{document}